\documentclass[12pt,nofootinbib,superscriptaddress]{revtex4}
\usepackage{amssymb,amsmath,latexsym,graphics, graphicx,epsfig}
\usepackage{latexsym}
\usepackage{amssymb}
\usepackage{epsfig,amsmath,graphics}
\usepackage{multirow}
\usepackage{appendix}
\usepackage{slashed}

\usepackage{color}
\usepackage{ulem}
\newcommand{\gsim}{\lower.7ex\hbox{$\;\stackrel{\textstyle>}{\sim}\;$}}
\newcommand{\lsim}{\lower.7ex\hbox{$\;\stackrel{\textstyle<}{\sim}\;$}}
\def\eg{{\it e.g.}}

\def\beq{\begin{equation}}
\def\eeq#1{\label{#1}\end{equation}}
\def\eeqn{\end{equation}}
\def\beqa{\begin{eqnarray}}
\def\eeqa#1{\label{#1}\end{eqnarray}}
\def\eeqan{\end{eqnarray}}

\def\leqn#1{(\ref{#1})}

\def\to{\rightarrow}
\def\Mltp{M_{\rm LTP}}

\begin{document}

%\begin{center}
%{\Huge \bf DRAFT - PRELIMINARY!!!}
%\end{center}
\vskip.5cm
\title{T-Quarks at the Large Hadron Collider: 2010-12}
\author{Maxim Perelstein}  
\email{mp325@cornell.edu}
\affiliation{Laboratory of Elementary Particle Physics, Physical Sciences Building, Cornell University, Ithaca, NY 14853, USA}
\author{Jing Shao} 
\email{jishao@syr.edu}
\affiliation{201 Physics Building, Syracuse University, Syracuse NY 13244, USA}

\vskip.5cm
\date{\today}
\begin{abstract}
We study the potential of the current Large Hadron Collider (LHC) 7 TeV run to search for heavy, colored vector-like fermions, which are assumed to carry a conserved $Z_2$ quantum number forcing them to be pair-produced. Each fermion is assumed to decay directly into a Standard Model quark and an invisible stable particle. T-odd quarks (T-quarks) and the lightest T-odd particle (LTP) of the Littlest Higgs model with T-parity provide an example of this setup. We estimate the bounds based on the published CMS search for events with jets and missing transverse energy in the 35 pb$^{-1}$ data set collected in the 2010 run. We find that T-quark masses below about 450 GeV are ruled out for the LTP mass about 100 GeV. This bound is somewhat stronger than the published Tevatron constraint. We also estimate the reach with higher integrated luminosities expected in the 2011-12 run. If no deviation from the SM is observed, we expect that a bound on the T-quark mass of about 650 GeV, for the LTP mass of 300 GeV and below, can be achieved with 1 fb$^{-1}$ of data. We comment on the possibility of using initial-state radiation jets to constrain the region with nearly-degenerate T-quark and LTP. 
\end{abstract}
 \maketitle

\newpage

\section{Introduction}

Searches for events with large missing transverse energy (MET) are a major focus of experiments at the Large Hadron Collider (LHC), motivated mainly by the fact that such events are generically predicted in supersymmetric extensions of the Standard Model (SM). The CMS and ATLAS collaborations recently published the results of the first such search, in the jets+MET channel, using the 35 pb$^{-1}$ of data collected during the 2010 run of the LHC with $\sqrt{s}=7$ TeV~\cite{CMS,ATLAS}. No excess over the expected SM background was observed, allowing to place bounds on the supersymmetric models which are already stronger than the previous Tevatron bounds. However, supersymmetry is not the only extension of the SM which predicts anomalous events with large MET. In fact, the prediction is very generic: all that's required of a model is that the heavy TeV-scale states carry a new conserved quantum number not carried by any SM particles, and that the lightest particle carrying this quantum number (which is automatically stable) be electrically neutral and color-singlet. Non-supersymmetric examples with these properties include models with universal extra dimensions (UED)~\cite{ued} and Little Higgs with T-parity (LHT)~\cite{LH,LHT,LHreview}. As in supersymmetry, the TeV-scale particles in these models can be paired with SM states of the same gauge quantum numbers; unlike supersymmetry, these theories predict that the heavy particles have the {\it same} spin as their SM partners. Searches for jets+MET events put constraints on these theories as well. In this paper, we will reinterpret the null result of the CMS search~\cite{CMS} in terms of bounds on a simple extension of the SM containing vector-like fermion partners for the light quarks, and an additional massive gauge boson, a partner of the SM photon (or, more precisely, the hypercharge gauge boson). This extension can be thought of as a ``simplified model"\footnote{See {\tt www.lhcnewphysics.org} for philosophy and examples of the simplified model approach.} corresponding to a limit of UED or LHT where all exotic particles other than the ones included in the model are either too heavy or too weakly coupled to be produced in significant numbers at the LHC (at least with current luminosity). In the LHT, such a situation is quite natural, and in fact this simplified version of the model has already been used as a basis for the Tevatron search for the T-odd quarks~\cite{LHT_TeV,LHT_D0}. We will thus refer to the heavy vector-like fermions as ``T-quarks". In UED, the Kaluza-Klein (KK) partners tend to be approximately degenerate in mass, and thus a simple truncation to just a handful of states is less likely to capture the phenomenology correctly; still, there is a limit of the model (which can be taken by adjusting brane-localized kinetic terms~\cite{ued2}) where our simple analysis would apply. 

Following Ref.~\cite{LHT_TeV}, we introduce four new vector-like (Dirac) fermions, $\tilde{Q}_i=(\tilde{U}_i, \tilde{D}_i)$, with the same gauge quantum numbers as the SM left-handed quarks $Q_i$, namely $({\bf 3}, {\bf 2})_{1/3}$. Here $i=1, 2$ is the generation index. There's also a discrete $Z_2$ symmetry, which we will call the T-parity, under which $\tilde{Q}_i\to-\tilde{Q}_i$. All SM states are invariant under this symmetry. We also introduce a massive vector boson $\tilde{B}^\mu$, which has no SM gauge charges, is odd under T-parity, and is coupled to fermions via
\beq
{\cal L}_{\rm int} = c g^\prime \left( \bar{Q}_i \gamma^\mu P_L \tilde{Q}_i + {~\rm h.c.}\right) \,B_\mu\,, 
\eeq{Lint}  
where sum over $i$ is implicit and $c$ is an order-one number whose exact value will play no role in our discussion, as long as the interaction is perturbative. We will assume that the four T-quarks have the same mass, $\tilde{M}$. This is motivated by constraints from flavor-changing neutral currents in the LHT model~\cite{LHTflavor}, as well as desire for maximal simplicity. We will further assume that $M(\tilde{B})<\tilde{M}$, so that the $\tilde{B}$ is the lightest T-odd particle (LTP), and is therefore stable. As discussed in~\cite{LHT_TeV}, this set of particles and mass hierarchies (with a few additional states above $\tilde{M}$ which will play no role in this analysis) arises naturally in the LHT model; moreover, $\tilde{B}$ can play the role of dark matter candidate~\cite{LHT_DM}, further motivating this spectrum.   

At the LHC, the dominant production mechanism in this model is pair-production of T-quarks via strong interactions:
\beq
q\bar{q}\to\tilde{U}_i\bar{\tilde{U}}_i,~~~gg\to\tilde{U}_i\bar{\tilde{U}}_i,
\eeq{process} 
and same for $\tilde{D}_i$. The produced T-quarks decay promptly via the interaction in Eq.~\leqn{Lint}, \eg 
\beq
\tilde{U}_i\to u_i \tilde{B},
\eeq{decay}
producing two high-$p_T$ jets and MET carried by the pair of $\tilde{B}$'s. 
Since no other decays are possible, the branching ratio of the decay~\leqn{decay} is one throughout the parameter space, independent of $c$. The phenomenology of the model is completely described by two parameters, $\tilde{M}$ and $\Mltp\equiv M(\tilde{B})$. The main goal of this paper is to estimate the current and near-future reach of the LHC search for jets+MET in this parameter space, taking the published CMS search~\cite{CMS} as the benchmark analysis. 

\section{Simulations, Analysis and MSSM Validation}

We implemented the model described above in {\tt MadGraph/MadEvent}~\cite{MG}.  The process~\leqn{process} was simulated using {\tt MadEvent}, and the decays~\leqn{decay} of the T-quarks were simulated using {\tt BRIDGE}~\cite{bridge}. (We checked, for a few representative points in the model parameter space, that the results are identical to simulating the production and decay together as a $2\rightarrow 4$ process in {\tt MadEvent}, but this approach is less computationally efficient and was not used in our scan over the model parameters.) The CTEQ6M PDF set~\cite{cteq} was used. Simulations were performed for a single T-quark flavor. We ignored the electroweak contribution to the production process, which is suppressed by a factor of order $\alpha^2/\alpha_s^2\sim 0.01$ with respect to the leading QCD contribution. In this approximation, production cross section and kinematic distributions are identical for the 4 T-quark flavors. The parton-level events created by {\tt MadEvent} were passed on to {\tt Pythia}~\cite{pythia} for hadronization and showering. A fast detector simulation was then performed using the {\tt PGS} (Pretty Good Simulation) package~\cite{PGS}, with the parameters set to approximate the CMS detector.\footnote{We used the file {\tt pgs\_card\_CMS.dat} included in the  
{\tt MadGraph/MadEvent} package to set the PGS parameters.} The resulting event sample was then subjected to the following set of cuts, which were chosen to approximate those used by the CMS study~\cite{CMS} as closely as possible:

\begin{enumerate}

\item Lepton veto: An event is rejected if a lepton (including an identified $\tau$) is present, with $p_T>10$ GeV, $|\eta|\leq 3$, and separated by $\Delta R\geq 0.4$ from every other object in the event (not including MET).

\item Jet acceptance: Jets with $p_T<50$ GeV or $|\eta|\geq 3$ are deleted. If the resulting event has $<2$ jets, it is discarded.

\item Jet selection: An event is accepted if and only if it has at least two jets with $p_T>100$ GeV, and the leading (highest-$p_T$) jet is at $|\eta|\leq 2.5$.

\item $H_T$ cut: An event is rejected if $H_T<350$ GeV, where $H_T$ is defined as the scalar sum of $p_T$'s of all jets in the event.

\item $\alpha_T$ cut: For a 2-jet event, we define $\alpha_T=E_{T2}/M_T$, where $E_{T2}$ is the energy of the least energetic of the jets, and $M_T$ is the transverse mass of the dijet system~\cite{alpha}. For an event with $\geq 3$ jets, we first combine the jets into 2 pseudo-jets, choosing the partition which minimizes the $E_T$ difference between the two pseudo-jets. We then define $\alpha_T$ in the same way as before, with pseudo-jets replacing the jets. In either case, the event is rejected if $\alpha_T<0.55$.   

\end{enumerate}  

We simulated 10000 events for each of 391 points in the LHT parameter space, spanning the ranges $\tilde{M}=220\ldots 700$ GeV, $\Mltp=20\ldots \tilde{M}-20$ GeV, with most of this area covered with a (20 GeV)x(20 GeV) grid. (We also explored additional points in the region of quasi-degenerate T-quark and LTP, see below.) For each point, we obtain the total cross section for production of the 4 flavors of T-quarks (which of course only depends on $\tilde{M}$), and the combined efficiency of the cuts. We then apply a K-factor to set the total cross section to its NLO value, with renormalization scale set to $\tilde{M}$~\cite{NLO_xsec}. (We used the simple parametrization of the NLO cross section given in the Appendix B of Ref.~\cite{berger}.)  Multiplying these numbers yields the total signal cross section after cuts, which can then be compared with present or extrapolated bounds inferred from the CMS analysis. 
We estimate that the combined renormalization scale and PDF uncertainty on the total cross section to be about 20\% (see, for example, Fig.~2~(c) in Ref.~\cite{berger}). By comparing cut efficiencies computed from simulations with fixed renormalization and factorization scales (which we use for our central-value estimates) to simulations with renormalization and factorization scales adjusted on event-by-event basis according to jet $p_T$'s, for a few representative points in the model parameter space, we estimate the Monte Carlo uncertainty in the efficiency calculation to be of order 10\%. Combining the two, we conservatively assign an uncertainty of 25\% to the theoretical prediction of the signal cross section after cuts.

\begin{table}[t]
\begin{center}
\begin{tabular}{|c||c|c||c|c|} \hline
                & \multicolumn{2}{|c||}{Total Efficiency} &  \multicolumn{2}{|c|}{Signature Efficiency} \\
\hline
channel & CMS & ~~~PS~~~ & CMS & PS \\
\hline
$\tilde{q}\tilde{q}$ & $16.0\pm0.1$ & 18.8 & $22.2\pm0.4$ & 21.7 \\
$\tilde{q}\tilde{g}$ & $14.4\pm0.1$ & 20.3 & $23.0\pm0.5$ & 23.9 \\
$\tilde{g}\tilde{g}$ & $12.0\pm0.4$ & 18.5 & $22.5\pm2.0$ & 22.6 \\
\hline
\end{tabular}
\caption{Signal efficiencies, in \%, for the cMSSM, LM1 point. ``CMS" denotes values reported by the CMS collaboration~\cite{CMS}; ``PS" refers to the analysis described in this section.}
\label{tab:valid}
\end{center}
\end{table}

To check that this simple procedure yields sensible results, we validated it by comparing the cut efficiencies for the supersymmetric signal at the point LM1 in the constrained minimal supersymmetric model (cMSSM), estimated using exactly the same procedure we use for T-quarks, with the corresponding efficiencies presented by CMS~\cite{CMS}. The results of this comparison are shown in Table~\ref{tab:valid}. Here, ``total" efficiencies denote the number of events passing all cuts normalized to the total number of events, while ``signature" efficiencies are normalized to the number of events after the lepton veto. It is clear that our procedure gives excellent match to the CMS results for the signature efficiencies, which are within 2 standard deviations for all 3 production channels. Our procedure does not work nearly as well for the total efficiencies. This is most likely due to our extremely simple-minded treatment of $\tau$'s, which are very common in the LM1 events since staus are light at that cMSSM point. Since the model we're studying in this paper does not predict any events with $\tau$'s (or indeed any SM leptons) at the parton level, we do not expect this issue to affect our results. Indeed, the lepton veto only removes 1-3\% of events in our T-quark event samples, so that even a factor of 2 error in the efficiency of this cut would only have a marginal effect on our results. On the other hand, the excellent agreement of our procedure with the CMS numbers at the level of signature efficiencies indicates that the procedure works very well when purely hadronic+MET events are considered. Since the vast majority of the T-quark events are of this kind, we expect our efficiencies for T-quark searches to be valid to a good approximation. (Of course, ideal agreement is not expected, since our modeling of the detector is quite crude.)

\section{Current and Projected Experimental Reach}

The sensitivity of an experiment to the new physics signal described above is limited by the SM backgrounds. Dominant SM sources of multijet+MET events are pure QCD events with mis-measured jet momenta, $Z$+jets events with $Z\to\nu\bar{\nu}$, $W$+jets events with $W\to \ell\nu$ and $\ell$ either not detected or misidentified, and $t\bar{t}$ events with one of the tops decaying leptonically. The CMS collaboration estimated the rate of SM background events passing the analysis cuts, by a combination of Monte Carlo simulations and data-based techniques. For example, an inclusive background estimate for the 35 pb$^{-1}$ data set considered in~\cite{CMS} is
\beq
{\rm Bg} = 9.4^{+4.8}_{-4.0}~{\rm (stat)}\pm1.0~{\rm (syst)}.
\eeq{bg_rate}
This number was obtained by measuring the rate of multijet+MET events in the ``control region", {\it i.e.} the region with lower $H_T$ than required in the analysis ($H_T\in [250, 350]$ GeV), and extrapolating to higher $H_T$. Alternative estimates for electroweak and $t\bar{t}$ processes were obtained by measuring related processes: for example, the $Z$+jets rate was estimated using the $\gamma$+jets sample. These estimates agree, within errors, with the inclusive estimate quoted above.    

The CMS collaboration observed 13 events passing the selection cuts in the data, consistent with the estimate~\leqn{bg_rate}. This measurement can be used to place a limit on the new physics contribution to the rate. A systematic derivation of such a limit should take into account the possibility of contamination of the control region by new physics. The procedure for doing so is rather complicated, and the result is model-dependent. On the other hand, given the shape of the SM $H_T$ distribution and the existing constraints on new physics models under consideration, the contamination is expected to be a minor effect, and at the level of phenomenological analysis of this paper, it is reasonable to ignore it. In this approximation, a new physics contribution after all cuts is limited to at most 13.4 events at 95\% confidence level (CL)~\cite{CMS}, corresponding to a cross section limit of 0.383 pb. 

\begin{figure}[t]
\centering%
\includegraphics[width=5in]{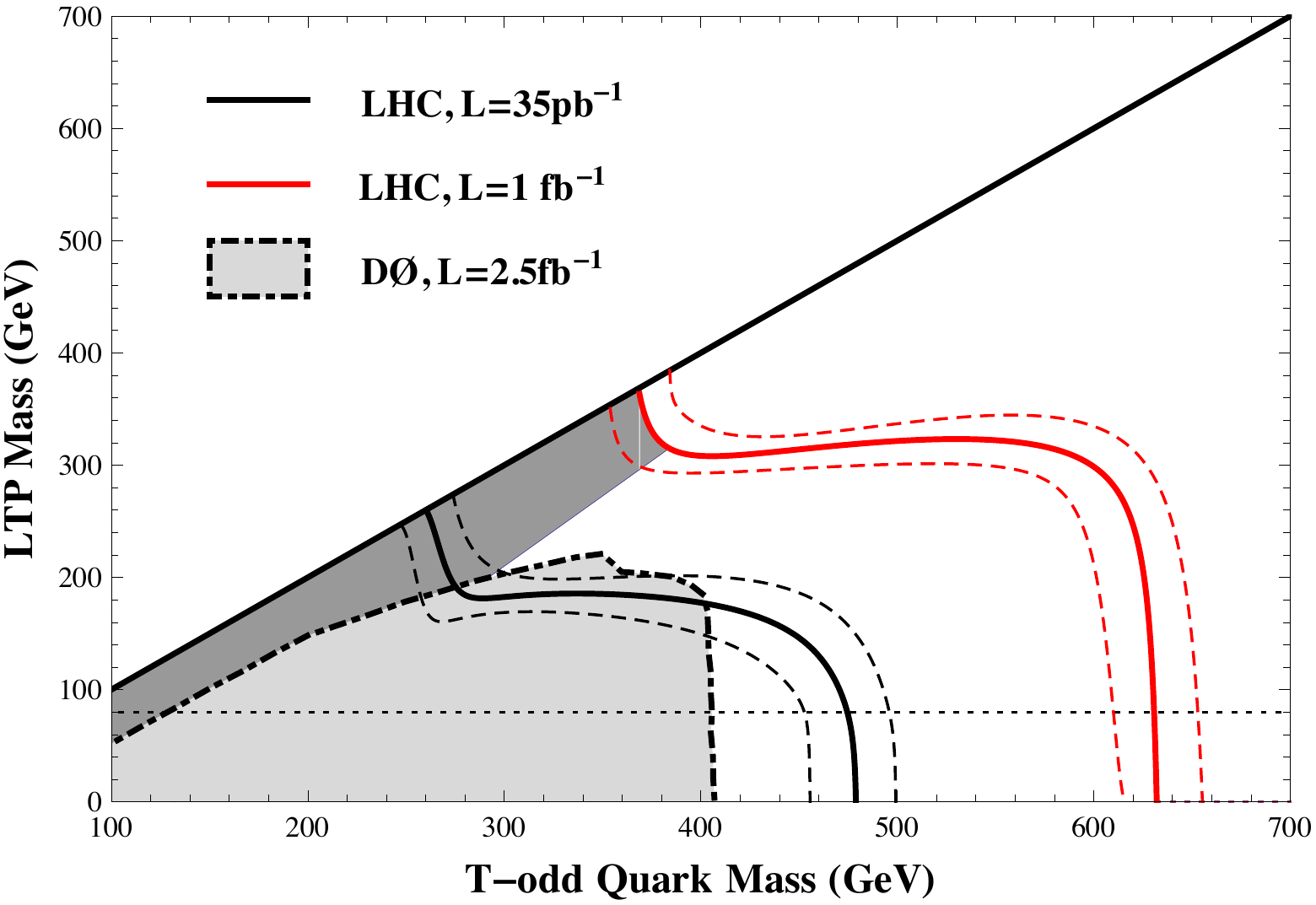}
\caption{Solid black line: Estimated exclusion contour based on the published CMS analysis~\cite{CMS}. Solid red line: Estimated reach for the same analysis with 1 fb$^{-1}$ of data at 7 TeV. Dashed black/red lines indicate the variation of the limits assuming a 25\% uncertainty on the cross section prediction. 
Lightly shaded parameter region below the dash-dotted black line is excluded by the D{\O} search at the Tevatron~\cite{LHT_D0}.  The region below the dotted line is ruled out by precision electroweak constraints in the LHT model~\cite{LHT_PEW}, but may be allowed in a more general simplified model context. In the darkly shaded band, the jets that pass the analysis cuts are predominantly from initial-state radiation.} 
\label{fig:excl}
\end{figure}

The impact of this cross section limit on the parameter space of our model is shown in Fig.~\ref{fig:excl}.   
The region to the left and below the solid black line is excluded. Notice that the bound is already stronger than the only published Tevatron bound on this model by the D{\O} collaboration~\cite{LHT_D0}, shown as the lightly shaded region on the figure. (To be fair, it should be noted that the D{\O} analysis was based on a 2.5 fb$^{-1}$ of the Tevatron data and has not been updated.) The LHC search sensitivity can be rapidly improved with larger data sets, expected to be collected during the 2011-12 run. As an example, we present an estimate of the reach expected with 1 fb$^{-1}$ of integrated luminosity (solid red line in Fig.~\ref{fig:excl}). To obtain this estimate, we rescaled the background rate from~\leqn{bg_rate}, assuming that the fractional statistical error will scale as $1/\sqrt{L_{\rm int}}$ while the fractional systematic error will remain unchanged at approximately $10$ \%. We further assumed that the measured rate will coincide with the central value of the background estimate. This procedure yields the expected cross section bound (after all cuts) of 74.6 fb, which was used to compute the reach. According to this simple extrapolation, the analysis becomes systematics-dominated around 1 fb$^{-1}$ of integrated luminosity, and further data does not significantly improve the reach: the expected cross section limit improves only to about 60 fb for 10 fb$^{-1}$.  

\begin{figure}[t]
\centering%
\includegraphics[width=3in]{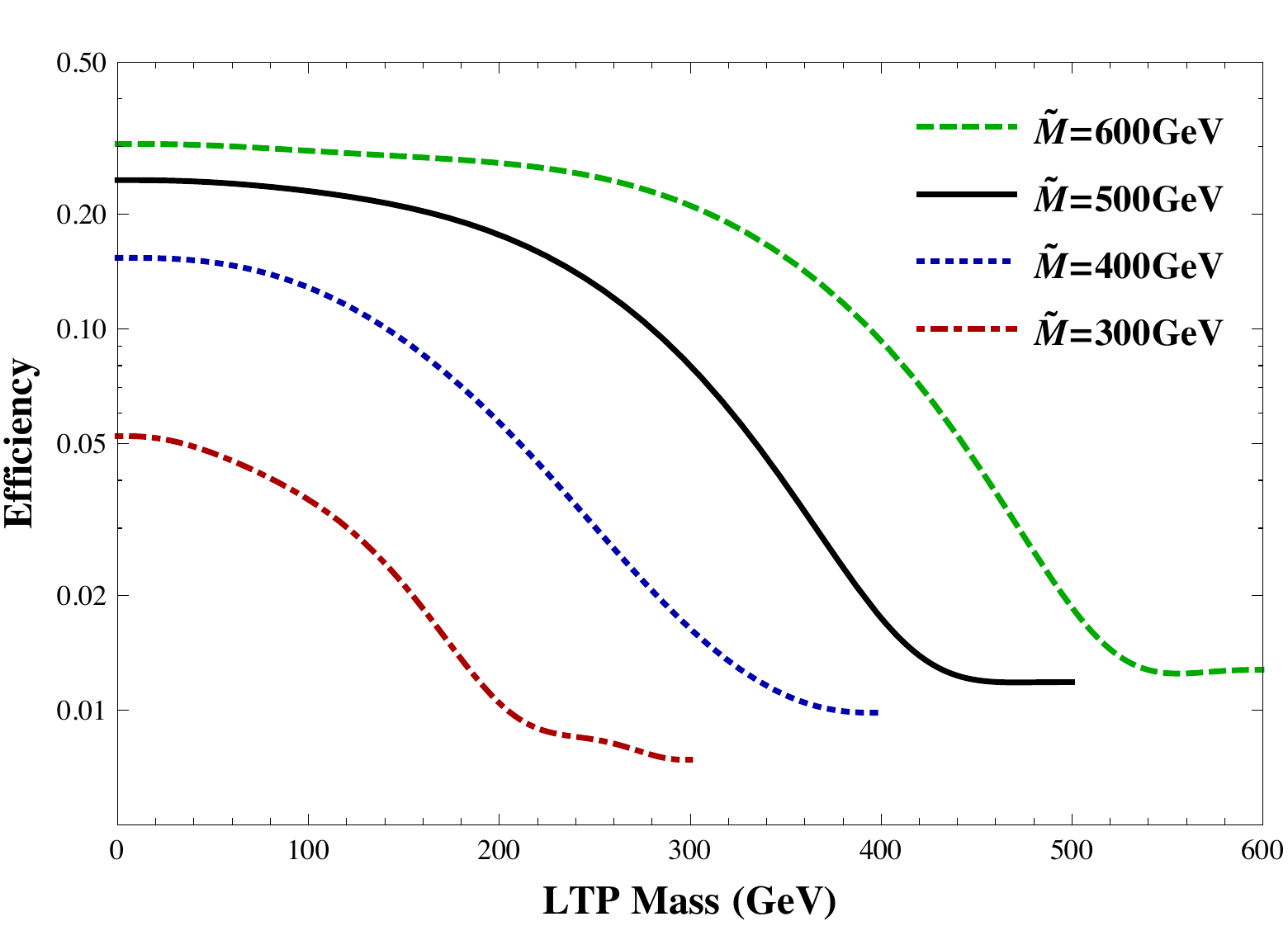}
\includegraphics[width=3in]{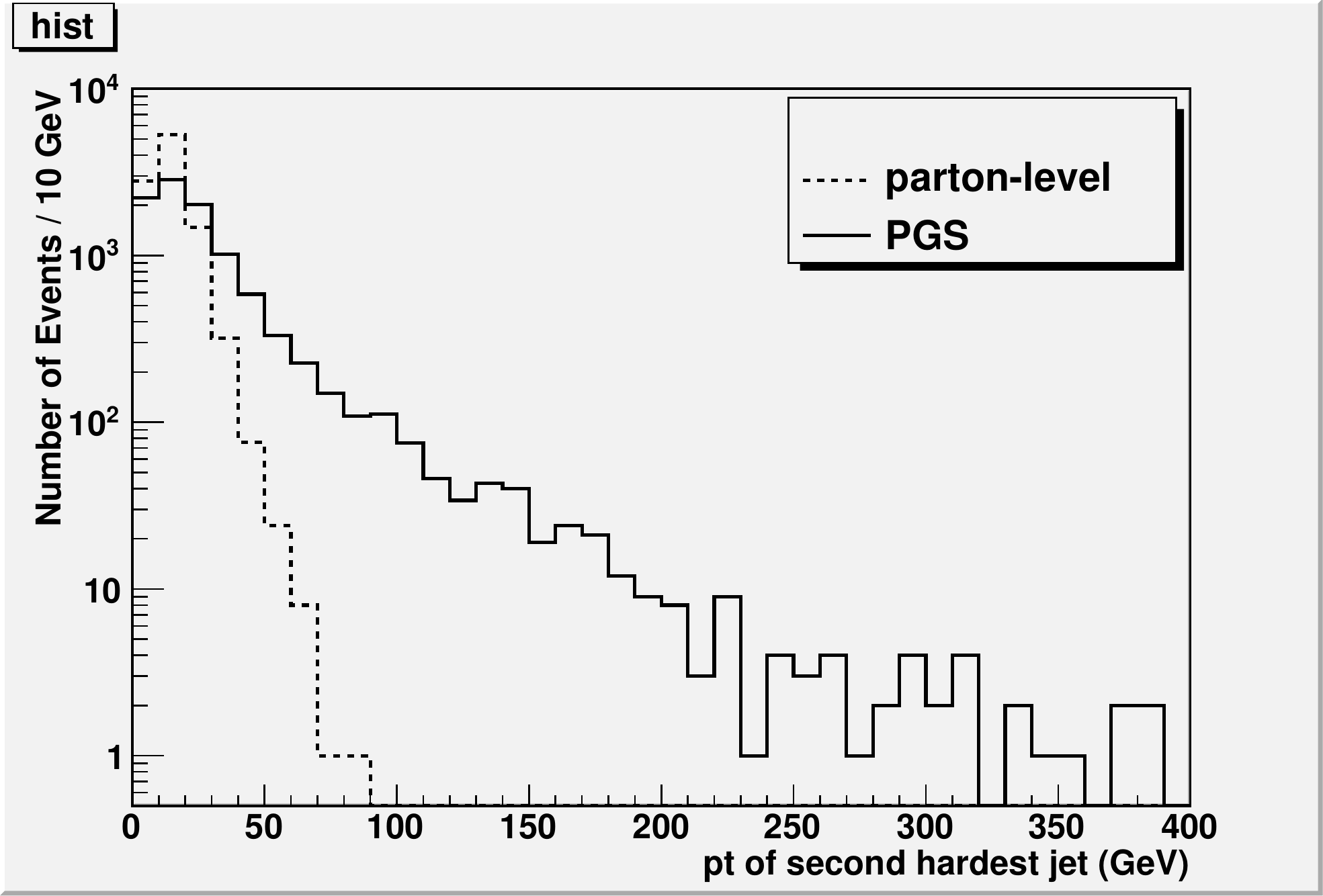}
\caption{Left: Combined analysis cut efficiency for the T-quark signal, as a function of the T-quark/LTP mass difference, for four sample values of the T-quark mass. Right: $p_T$ distribution of the second-hardest jet in each event, at the parton level (dotted histogram) and at the level of {\tt PGS} output (solid histogram). In this simulation, $\tilde{M}=400$ GeV and $\Mltp=380$ GeV.} 
\label{fig:eff}
\end{figure}

A somewhat surprising feature of Fig.~\ref{fig:excl} is the apparent sensitivity of the analysis in the region of quasi-degenerate T-quark and LTP. Naively, the cut efficiency should become vanishingly small in that region: The jet energy in the T-quark rest frame is proportional to $\tilde{M}-\Mltp$, and the T-quarks are not ultrarelativistic in the lab frame (the typical velocity is about 0.5), so the jets should be soft and should fail the acceptance and $H_T$ cuts. In fact, the cut efficiency decreases as expected for  $\tilde{M}-\Mltp\gsim 100$ GeV, but then approximately flattens out (at about 1\% for the typical T-quark mass in our analysis) at $\tilde{M}-\Mltp \lsim 100$ GeV (see the left panel of Fig.~\ref{fig:eff}). Our interpretation is that in this region, the hard jets passing the acceptance and $H_T$ cuts are due to initial-state radiation (ISR). This is further confirmed by comparing parton-level and {\tt Pythia}-level events for a sample parameter point in the quasi-degenerate region, shown on the right panel of Fig.~\ref{fig:eff}. The parton-level sample has no jets passing acceptance cuts, while {\tt Pythia}-level $p_T$ distribution has a long tail extending above the 100 GeV threshold required in this analysis. Of course, the treatment of ISR in {\tt Pythia} is approximate, and large corrections are expected at large $p_T$. In our events the typical ISR jet passing the cut has $p_T\sim 100$ GeV, which is hierarchically smaller than the typical hard scale of the T-quark production process 
$\sqrt{\hat{s}}\sim 2\tilde{M}\sim 1$ TeV, so the {\tt Pythia} predictions should not be grossly incorrect. Still, a more detailed treatment (with careful jet matching, or, ideally, with fully differential NLO cross sections in addition to resummation of large logs) is needed to obtain fully reliable efficiency estimates in the quasi-degenerate region. To emphasize this point, in Fig.~\ref{fig:excl} we shaded the region where most of the jets passing the analysis cuts are from ISR. 

\section{Conclusions}

In this paper, we estimated the current and near-future LHC sensitivity to ``T-quarks", exotic vector-like quarks carrying a $Z_2$ charge, which decay to an ``LTP", a weakly-interacting massive particle invisible in the detector, and an SM quark. We used the published CMS search in the jets+MET channel~\cite{CMS} as the benchmark analysis. We found that this search, based on the 35 pb$^{-1}$ of data at $\sqrt{s}=7$ TeV collected in 2010, already improves on the existing Tevatron bounds. Significant future improvement can be obtained with 1 fb$^{-1}$ data set, expected to be collected in 2011-12. Beyond 1 fb$^{-1}$, we find that the analysis becomes systematics-limited, assuming that the fractional systematic error on the background estimate reported by CMS (about 10\%) does not improve with more data.

While we used the terminology of the Little Higgs model with T-parity (LTP) to describe the new particles present in our model, the analysis is in fact quite model-independent, and can be applied to any new physics model whose LHC phenomenology can be approximated by the simple Lagrangian in~\leqn{Lint}.
In this sense, our study is in the spirit of the ``simplified model" approach to describing LHC signals of new physics. In particular, our analysis applies (at least approximately) to the UED model, in the limit when KK partner of the gluon is significantly heavier than the KK quarks.

An interesting feature that was noted is that in the region of nearly-degenerate T-quark and LTP, the cut efficiencies do not vanish, due to the presence of ISR jets passing the acceptance and analysis cuts in a non-vanishing fraction of events. This can be especially important in the UED context, where such near-degenerate spectrum is generic (at least in the absence of large brane-localized kinetic terms). A more careful analysis of ISR jets is needed in this region. (See Ref.~\cite{Jay} for similar observations in the context of supersymmetry searches.)

On a more technical note, we found that, reassuringly, the Monte Carlo tools widely used by phenomenologists, including {\tt PGS}, do a decent job of reproducing experimental efficiencies for fully hadronic events. However, we did not find a good agreement with CMS lepton veto efficiencies, most likely due to our overly simplistic treatment of $\tau$'s. This was not an issue in our analysis, since our signal model predicted essentially no leptons, but it does indicate that caution is warranted in general.

To conclude, our analysis shows that the 7 TeV LHC runs in 2010-2012 have a potential to either discover the T-quarks or substantially improve the existing bounds. We encourage the LHC experiments to perform this search. If an excess above the SM background is observed in this channel, we stress that alternative interpretations, such as the model studied here, must be considered before it is attributed to supersymmetry. (For a recent analysis of model discrimination power of the LHC in this channel based on jet $p_T$ and angular distributions, see Ref.~\cite{CMS_MD}.) 

\section{Acknowledgements}

We are grateful to Jay Hubisz for many useful discussions, and to David Curtin for help with software. MP is supported by the U.S. National Science Foundation through grant PHY-0757868 and CAREER award PHY-0844667. JS is supported by the Syracuse University College of Arts and Sciences. 

%%%%%%%%%%%%%%%%%%%%%%%%%%%%%

\end{document}